\documentclass{ws-p10x7}
\usepackage{graphicx}

\begin{document}

\title{The Hadronic Picture of the Photon}

\author{Thorsten Wengler}

\address{CERN, EP division, CH-1211 Geneve 23, Switzerland\\
         email: Thorsten.Wengler@cern.ch}

\twocolumn[\maketitle\abstract{
The hadronic interactions of the photon are studied in terms of new
measurements of the total hadronic cross section and di-jet production
in photon-photon collisions at LEP2.}]

\section{Total hadronic {$\gamma\gamma$} cross section}

The reaction of
${e^+e^-}$ ${\rightarrow}$ ${e^+e^-}$ ${\gamma^*\gamma^*}$ 
${\rightarrow}$ ${e^+e^-}+{hadrons}$ is analysed for quasi-real photons
using data collected by L3 and OPAL at LEP2. From this measurement the
total hadronic cross section $\sigma_{\gamma\gamma}(W)$ is extracted
using a luminosity function for the photon flux and extrapolating to 
$Q^2$=0~GeV$^2$. The L3 data~\cite{l3xstot} are shown in 
Figure~\ref{fig:xstot}, representing an integrated luminosity of
392.6~pb$^{-1}$ taken at $\sqrt{s}$ from 189 to 202 GeV. The total
hadronic cross section, $\sigma_{\gamma\gamma}$, is shown as a
function of the invariant mass $W$ of the photon-photon system. Also
shown as the solid line is the result of a fit using the
parameterisation proposed by Donnachie and
Landshoff~\cite{donlan}: 
$\sigma_{\mathrm{tot}} = A s^{\epsilon} + B s^{-\eta}$. The second
term represents the Reggeon exchange dominant at low $W$, while
at high $W$ the Pomeron exchange driven by the exponent $\epsilon$
prevails. If photons behave predominantly like hadrons the values of 
$\epsilon=$ 0.095$\pm$ 0.002 and $\eta=$ 0.34$\pm$ 0.02 obtained from
a universal fit to hadron-hadron cross sections~\cite{pdg} should be
valid for $\sigma_{\gamma\gamma}(W)$ as well. Indeed this is not the
case for the L3 data, as demonstrated by the dashed line in 
Figure~\ref{fig:xstot}. Instead a fit with A, B, and $\epsilon$ as
free parameters yields $\epsilon=$ 0.250$\pm$0.016, which is more than
a factor of two higher than the universal value, a rise significantly
steeper 
than expected from the universal fit of hadron-hadron cross
sections. The 
L3 data is compared to an OPAL measurement~\cite{opxstot} using an
integrated 
luminosity of 74.3~pb$^{-1}$ taken at $\sqrt{s}$ from 161 to
183~GeV. Both measurements are consistent inside their respective
uncertainties. A similar fit as described above to the OPAL data
however yields $\epsilon=$ 0.101$\pm$ 0.025, in agreement with the
universal fit. The steeper rise observed by L3 is mainly determined by
their data points at highest and lowest $W$.
\begin{figure}[ht]
\begin{center}
\includegraphics[width=0.80\columnwidth]{fig1.eps}
\end{center}
\caption{The photon-photon total hadronic cross section, 
$\sigma_{\gamma\gamma}(W)$. The solid line is a fit to the L3 data,
the dashed line uses the 
parameters from a universal fit to hadron-hadron cross sections (see
text). Here the L3 points do not 
include the dominant systematic error due to MC model
dependencies. With this error included the total uncertainties on the OPAL
and L3 measurements are of comparable size.}
\label{fig:xstot}
\end{figure}

\vspace*{-0.5cm}
\section{Di-jet Production}
 \begin{figure}[ht]
\begin{center}
\includegraphics[width=0.8\columnwidth]{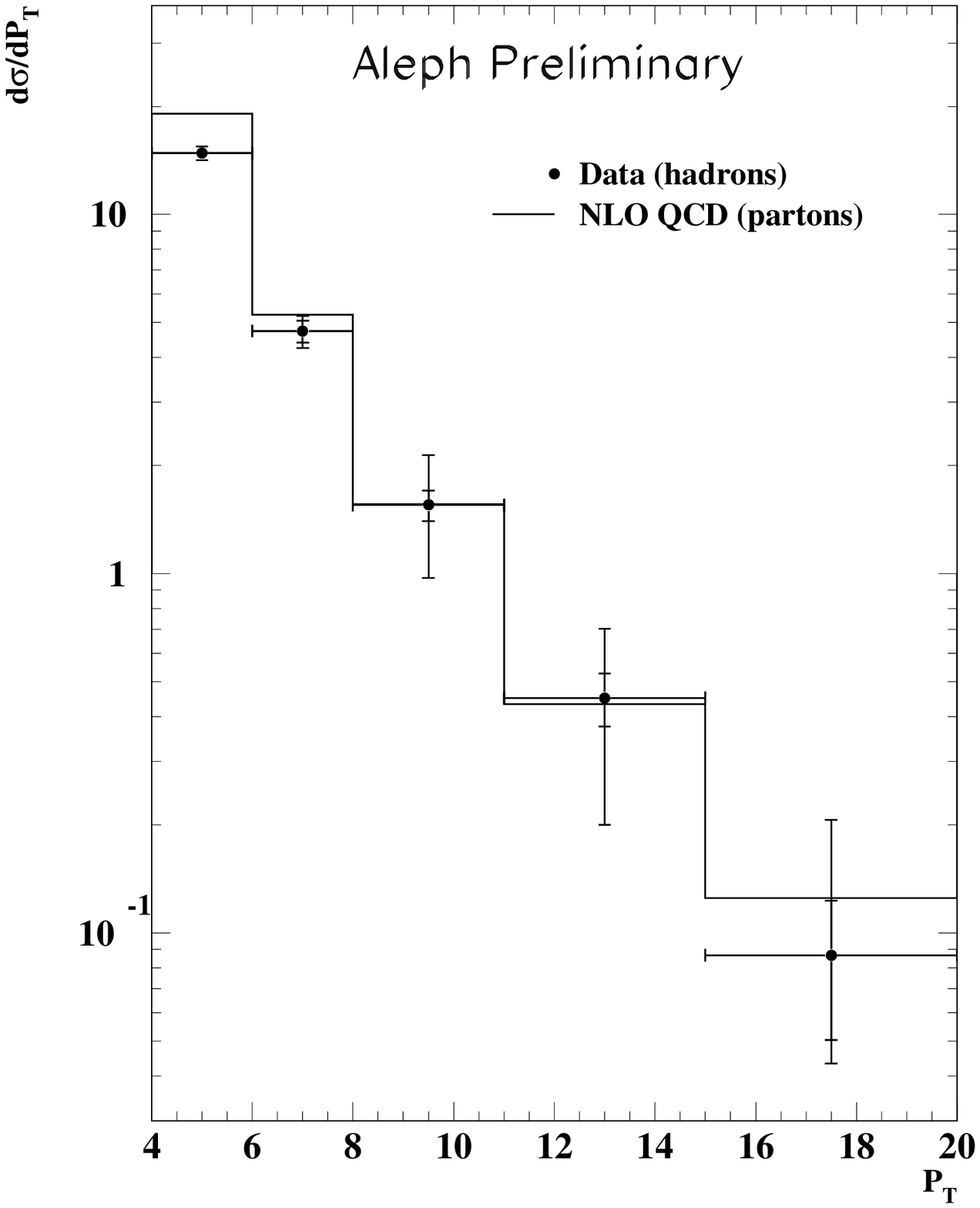}
\end{center}
\caption{Differential di-jet cross section as measured by ALEPH using
59.2~pb$^{-1}$ taken at $\sqrt{s}$ =183~GeV.}
\label{fig:djal}
\end{figure}

The production of di-jets in the collisions of two quasi-real photons
has been measured by ALEPH and OPAL. Jet production is calculable in
perturbative QCD. The measurements can therefore be used to study the
performance of NLO predictions for these processes. 
ALEPH~\cite{aldjet} has analysed
59.2~pb$^{-1}$ taken at $\sqrt{s}$ =183~GeV using the
Durham $k_\perp$ clustering 
algorithm~\cite{ktclus} with  $Y_{\mathrm{cut}}=$0.018, chosen to
maximise the rate of di-jet events. Di-jet events are required to have
at least two jets with $|\eta_{\mathrm{jet}}|<1.4$, and a minimum
jet transverse momentum of 4 (3)~GeV for the first (second) jet.
The differential di-jet cross section as a function of the highest
$p_T$ jet in the event is shown in Figure~\ref{fig:djal}. The data is
compared to an NLO calculation of Klasen et al.\cite{klasen},
performed at the level of QCD partons. The agreement of the
calculation with the data is good, except 
at the lowest $p_T^{\mathrm{jet}}$.

\begin{table}[b]
\begin{center}
\begin{tabular}{|l|c|c|}
\hline
&data $\left[\mathrm{pb}\right]$ & NLO $\left[\mathrm{pb}\right]$\\
\hline
\multicolumn{3}{|l|}{$5~\mathrm{GeV} < 
\bar{E}_{\mathrm{T}}^{\mathrm{jet}} < 7~\mathrm{GeV}$}\\
\hline
$x_\gamma >0.75$ & 111.0$\pm$3.8 & 206.5 \\
 $x_\gamma <0.75$ & 205.5$\pm$4.8 & 84.4 \\
\hline
\multicolumn{3}{|l|}{$7~\mathrm{GeV} <
  \bar{E}_{\mathrm{T}}^{\mathrm{jet}} < 11~\mathrm{GeV}$}\\ 
\hline
$x_\gamma >0.75$ & 77.4$\pm$2.6 & 87.0 \\
$x_\gamma <0.75$ & 71.5$\pm$2.2 & 32.6 \\
\hline
\multicolumn{3}{|l|}{$11~\mathrm{GeV} <
  \bar{E}_{\mathrm{T}}^{\mathrm{jet}} < 25~\mathrm{GeV}$}\\ 
\hline
$x_\gamma >0.75$ & 32.0$\pm$2.5 & 27.4 \\
$x_\gamma <0.75$ & 15.8$\pm$1.7 & 7.5 \\
\hline
\end{tabular}
\end{center}
\caption{OPAL di-jet cross sections compared to NLO predictions in
  regions of $x_\gamma$ and $\bar{E}_{\mathrm{T}}^{\mathrm{jet}}$.}
\label{nlotab}
\end{table}

\begin{figure*}[t]
\epsfxsize24pc
\figurebox{}{}{fig3.eps}
\caption{OPAL differential di-jet cross section as a function of $x_\gamma$
  in several regions of $\bar{E}_{\mathrm{T}}^{\mathrm{jet}}$.} 
\label{fig:djopxg}
\end{figure*}

OPAL~\cite{opdjet} has analysed 384~pb$^{-1}$ taken at $\sqrt{s}$ from
189~GeV to 202~GeV. Jets are found using the inclusive
$k_\perp$~\cite{ktclus} clustering algorithm. Jets entering the
analysis have a
pseudorapidity  $|\eta_{\mathrm{jet}}|<2.0$. The two jets with the
highest $E_{\mathrm{T}}^{\mathrm{jet}}$ in each event 
are taken. An average transverse energy of the two leading jets in the
event of $\bar{E}_{\mathrm{T}}^{\mathrm{jet}}>$ 5~GeV is required. The
additional condition $|E_{\mathrm{T,1}}^{\mathrm{jet}} -
E_{\mathrm{T,2}}^{\mathrm{jet}}| / (E_{\mathrm{T,1}}^{\mathrm{jet}} +
E_{\mathrm{T,2}}^{\mathrm{jet}}) < 1/4$ keeps low
$E_{\mathrm{T}}^{\mathrm{jet}}$ jets from entering the distributions
and ensures asymmetric $E_{\mathrm{T}}^{\mathrm{jet}}$ thresholds for
the two jets. The differential cross section as a function of 
$\bar{E}_{\mathrm{T}}^{\mathrm{jet}}$ (not shown) is reasonably well
described by both the LO MC
models PHOJET and PYTHIA and the NLO calculation~\cite{klasen}. Di-jet
events have the 
particular advantage that the fraction of the photon momentum,
$x_\gamma$, entering the hard scattering can be estimated from the
di-jet system. The observable quantity 
$x_{\gamma}^{\pm} \equiv (\sum_{\rm jets=1,2}(E{\pm}p_z))/
(\sum_{\rm hadrons}(E{\pm}p_z))$ is defined for this purpose.
OPAL has for the first time measured differential cross sections as a
function of $x_\gamma$, where the data has been fully unfolded for
detector effects. Here $x_\gamma$ refers to both $x_{\gamma}^{+}$ and 
$x_{\gamma}^{-}$ entering the distribution. The region of small
$x_\gamma$ is expected to be dominated by resolved photon
interactions, and is hence particularly sensitive to the gluon density
in the photon. The OPAL data is shown in
Figure \ref{fig:djopxg} in three regions of
$\bar{E}_{\mathrm{T}}^{\mathrm{jet}}$. 
The data is compared to three
predictions of the PHOJET generator using the GRV LO, SaS1D, and LAC1
parton distribution functions respectively. The sensitivity of this
observable to the different parton distributions is clearly
visible. The already disfavoured LAC1 set is shown to demonstrate the
effect of a high gluon density and clearly overshoots the data
especially at low $\bar{E}_{\mathrm{T}}^{\mathrm{jet}}$. GRV and SaS1D
on the other hand seem to underestimated the gluon density. A
preliminary study of the underlying event using PYTHIA with and
without multiple interactions show a visible effect only at the lowest
values of $\bar{E}_{\mathrm{T}}^{\mathrm{jet}}$ and
$x_\gamma$. Table~\ref{nlotab} compares the OPAL measurement with an
NLO calculation~\cite{klasen} for small and large $x_\gamma$ in the
three regions of  
$\bar{E}_{\mathrm{T}}^{\mathrm{jet}}$. It should be stressed
that hadronisation corrections are not taken into account in this
calculation. 
The NLO calculation predicts much too
large a cross section for $x_\gamma>0.75$ and small
$\bar{E}_{\mathrm{T}}^{\mathrm{jet}}$, while being 
too low by about a factor of two for $x_\gamma<0.75$. With increasing
$\bar{E}_{\mathrm{T}}^{\mathrm{jet}}$ the discrepancy for $x_\gamma>0.75$
largely disappears, while for 
$x_\gamma<0.75$ the NLO prediction remains too low by a factor of two for
the highest $x_\gamma>0.75$ considered. This also suggests that the parton
density functions used in the NLO calculation (GRV) underestimate the gluon
density in the photon.

\end{document}